\documentstyle[12pt,epsf]{article}
\newcommand {\beq}{\begin{equation}}
\newcommand {\eeq}{\end{equation}}
\newcommand {\beqa}{\begin{eqnarray}}
\newcommand {\eeqa}{\end{eqnarray}}
\newcommand {\n}{\nonumber \\}
\def\ep{\epsilon}
\def\th{\theta}
\def\la{\lambda}
\def\da{\dot{a}}
\def\pa{\partial}

\begin{document}
\setlength{\oddsidemargin}{0cm}
\setlength{\baselineskip}{7mm}

\begin{titlepage}
 \renewcommand{\thefootnote}{\fnsymbol{footnote}}
$\mbox{ }$
\begin{flushright}
\begin{tabular}{l}
KEK-TH-518 \\
TIT/HEP-369\\
May 1997
\end{tabular}
\end{flushright}

~~\\
~~\\
~~\\

\vspace*{0cm}
    \begin{Large}
       \vspace{2cm}
       \begin{center}
         {String Field Theory from IIB Matrix Model}      \\
       \end{center}
    \end{Large}

  \vspace{1cm}

\begin{center}
           Masafumi F{\sc ukuma}$^{1)}$\footnote
            {
e-mail address : fukuma@yukawa.kyoto-u.ac.jp},
           Hikaru K{\sc awai}$^{2)}$\footnote
           {
e-mail address : kawaih@theory.kek.jp},
           Yoshihisa K{\sc itazawa}$^{3)}$\footnote
           {
e-mail address : kitazawa@th.phys.titech.ac.jp}{\sc and}
           Asato T{\sc suchiya}$^{2)}$\footnote
           {e-mail address : tsuchiya@theory.kek.jp,{}~JSPS Research Fellow}\\
      \vspace{1cm}
        $^{1)}$ {\it Yukawa Institute for Theoretical Physics,
Kyoto University, }\\
                 {\it Kitashirakawa Sakyo-ku, Kyoto 606-01, Japan}\\

        $^{2)}$ {\it National Laboratory for High Energy Physics (KEK),}\\
               {\it Tsukuba, Ibaraki 305, Japan} \\
        $^{3)}$ {\it Department of Physics, 
Tokyo Institute of Technology,} \\
                 {\it Oh-okayama, Meguro-ku, Tokyo 152, Japan}\\
\end{center}

\vfill

\begin{abstract}
\noindent 
We derive Schwinger-Dyson equations for the Wilson loops of a type IIB
matrix model. Superstring coordinates are introduced through the construction 
of the loop space.
We show that the continuum limit of the loop equation
reproduces the light-cone superstring field theory of type IIB superstring
in the
large-N limit. We find that the interacting string theory
can be obtained in the double scaling limit as it is expected. 
\end{abstract}
\vfill
\end{titlepage}
\vfil\eject

\section{Introduction}
\setcounter{equation}{0}
The recent development in our understanding of nonperturbative
aspects of string theory has reinforced the connection between string theory
and gauge theories\cite{MS}\cite{Witten}\cite{Polchinski}.  
It has been well known that the gauge theories
can be obtained in the low energy limit of string theory.
On the other hand the nonperturbative reformulation of superstring theory
appears to be possible now in terms of the D-particles\cite{BFSS} or the
D-instantons
\cite{IKKT}.  They are nothing but large-N (partially) reduced models of ten 
dimensional super Yang-Mills theory.

It has been long hoped that the gauge theories may be solvable in the large-N
limit as a kind of string theory\cite{thoot}. 
In gauge theories, the Wilson loops are the natural candidates
to be identified with strings.  They indeed obey the loop equations which
resemble
the Virasoro constraints in string theory.
Hence the loop equations have been investigated
in the hope of deriving the (light-cone) Hamiltonian of string
theory\cite{makeenko}.
The large-N reduced models have been originally invented in such a
context\cite{RM}
and the relation to bosonic string theory has been noted\cite{Bars}\cite{Zacos}.
These hopes appear to be realizable now for ten dimensional super Yang-Mills
theory in a most dramatic setting which involves the theory of gravitation
and the unification of forces. 

Our matrix model for type IIB superstring has the manifest $N=2$
super Poincare invariance in ten dimensions. It has the lowest dimensional
constituents (D-instantons) and the highest symmetry of this class of the
matrix models.
We hope it will be most useful to investigate the nonperturbative aspects of superstring theory.
The related models and the possible extensions have also been proposed
\cite{periwal}\cite{Li}\cite{chepelev}\cite{smith}\cite{olesen}\cite{yoneya}\cite{olesen2}.
Qualitatively reasonable results have already been obtained at long distance 
by the matrix models
when we investigate the interactions between the  D-strings at the one loop 
level. However the standard folklore suggests that 
we may need to sum up all planar diagrams of the matrix models 
even to obtain free strings.
Therefore we need to solve the theory in the large-N limit. 
Since we know a possible solution (superstring) to this question, 
this task is not inconceivable.
Our aim of this paper is in fact to derive the light-cone field theory of 
Green, Schwarz and Brink\cite{BGS} for type IIB superstring 
from a matrix model\cite{IKKT}. 

The only successful string field theories so far are formulated in the
light-cone gauge.  In this respect they may not be effective to study
nonperturbative aspects of string theory. However our motivation 
to derive the light-cone superstring field theory from the matrix model here
is to prove that the matrix model reproduces the standard string
perturbation theory.

In section 2, we introduce the Wilson loops as the operators
which create and annihilate strings (string fields).  The string coordinates are
introduced through the Wilson loops. We then discuss the continuum limit of
the differential
operators in the loop space.
In section 3, we set up the loop equations for the Wilson loops
by using the Schwinger-Dyson equations. We demonstrate that the loop equations
reduce to that of free supersrting Hamiltonian in the large-N limit. 
We also explain that the structure of the cubic light-cone Hamiltonian of
superstring naturally
appears from the matrix model.
In section 4, we construct a general proof of our assertion that the IIB
matrix model
reproduces the light-cone superstring field theory in the double scaling limit.
We construct supercharges of the matrix model in the loop space
and show that they agree with those of superstring theory.
The proof is based on the power counting
and the symmetry arguments.
We conclude in section 5 with discussions.

One of the crucial issues of the matrix model approach is to reproduce the
standard string
perturbation theory. We are successful in this respect since we have derived
light-cone Hamiltonian of superstring from the matrix model. Therefore we have
proven our previous conjecture that our matrix model is indeed a
nonperturbative 
formulation of type IIB superstring theory.

\section{Strings and the Wilson loops}
\setcounter{equation}{0}
We recall our matrix model action\cite{IKKT} which is a large-N reduced model
of ten dimensional super Yang-Mills theory:
\beq
S  =  -{1\over g^2}Tr({1\over 4}[A_{\mu},A_{\nu}][A^{\mu},A^{\nu}]
+{1\over 2}\bar{\psi}\Gamma ^{\mu}[A_{\mu},\psi ]) ,
\label{action}
\eeq
where $\psi$ is a ten dimensional Majorana-Weyl spinor field, and 
$A_{\mu}$ and $\psi$ are $N \times N$ Hermitian matrices.
We have interpreted some of the classical solutions of this action as D-strings
and shown that they indeed
interact in a consistent way with this interpretation at long distance.

We have also proposed that the Wilson loops are the creation and annihilation
operators for
strings\cite{IKKT}. 
The Wilson loops can wind many times around the world even
in the reduced models which have only few (even single) points 
in the universe. If we consider the  T-duality, the winding numbers can be
reinterpreted as the momentum.  
Based on these reasonings, we consider the following regularized Wilson loops:
\beqa
w(C) & = & Tr[v(C)]  ,\n
v(C) &=&
\prod_{n=1}^M U_n ,\n
U_n &=&exp\{i\epsilon (k_{n}^{\mu}A_{\mu}
+\bar{\lambda }_{n}\psi )\} .
\label{Wilsonloop}
\eeqa
Here $k_n^{\mu}$ are momentum densities distributed along $C$, and 
we have also introduced the fermionic sources 
$\lambda _{n}$ .
$\epsilon$ in the argument of the exponential 
is a cutoff factor.
As we will see in the subsequent sections, $\ep$ can be regarded as a lattice
spacing of the worldsheet. We therefore call $\ep \rightarrow 0$ limit
the continuum limit. 
We may also  Fourier transform the Wilson loop from the momentum space 
$k^{\mu}_n$ to the real space $x^{\mu}_n$.

In the next section, we consider the continuum limit of the Wilson loops.
As a preparation we study the differential operators in the 
loop space and the operator insertions into the Wilson loops
in the remaining part of this section . 
Let us consider the operator insertion of $A_{\mu}$ between the $n-$th 
and the $(n+1)-$th links:
\beq
U_n A_{\mu} U_{n+1}.
\label{opein0}
\eeq
To the leading order of $\ep$, it is equivalent to a simple differential
operator in the loop space:
\beqa
&&{\pa\over \pa i\ep k^{\mu}_{n}} U_n U_{n+1}\n
&=& 
U_n \int _0^1 dt exp\{-i\ep t (k^{\nu}_n A_{\nu} +\bar{\lambda}_n\psi )\}
A_{\mu} exp\{i\ep t (k^{\nu}_n A_{\nu} +\bar{\lambda}_n\psi )\}U_{n+1}\n
&=&U_n
(A_{\mu}-{i\ep\over 2}[k^{\nu}_n A_{\nu} +\bar{\lambda}_n\psi,A_{\mu}]
+\cdots)U_{n+1}.
\label{opein}
\eeqa
However eqs.(\ref{opein0}) and (\ref{opein}) differ from each other 
at the next order of $\ep$ by the commutators. 
The commutators such as $[A_{\mu},A_{\nu}]$
can be expressed by differential operators in the loop space as follows:
\beqa
&&({\pa\over \pa i\ep k^{\mu}_{n}} -{\pa\over \pa i\ep k^{\mu}_{n+1}})
({\pa\over \pa i\ep k^{\nu}_{n}} +{\pa\over \pa i\ep k^{\nu}_{n+1}})
U_n U_{n+1} \n
&=&
U_n([A_{\mu},A_{\nu}]
+\cdots )U_{n+1} .
\eeqa
Therefore the leading commutators in eq.(\ref{opein}) can be expressed
in terms of differential operators in the loops space. 
So the operator insertion (\ref{opein0}) can be expressed by a
differential operator
in the loop space up to $O(\ep )$. The remaining commutators which involve
three matrix
fields can be expressed by differential operators in an analogous way
if we consider three neighboring links. In this way in principle 
any operator
insertion
into the Wilson loop can be expressed by an infinite series of 
local differential operators in the loop space. 

We also consider the differentiation of the Wilson loops by
the matrix valued fields. 
The differentiation of the link variable by $A_{\mu}$
is given by.
\beqa
&&{\pa \over \pa A^{\alpha}_{\mu}} U_n
\n
&=&
i\ep k^{\mu}_n\int _0^1 dt exp\{i\ep t(k^{\nu}_n A_{\nu}
+\bar{\lambda}_n\psi )\}t^{\alpha}
exp\{-i\ep t(k^{\rho}_n A_{\rho} +\bar{\lambda}_n\psi )\}
U_n \n
&=&
i\ep k^{\mu}_n(t^{\alpha} +{i\ep\over 2} [k^{\nu}_n A_{\nu}
+\bar{\lambda}_n\psi ,t^{\alpha}]
 +\cdots )
U_n .
\eeqa
Here $t^{\alpha}$ denotes the generators of the gauge group $U(N)$.
Obviously the leading term is  
the multiplication by $i\ep k^{\mu}_n t^{\alpha}$. Again the remaining
terms can be expressed by some infinite series of differential operators.

Finally we make a comment on the universality of the differential operators
in the loop space.
What we are interested in is the continuum limit such as
\beq
y_{n+1}-y_n
\sim \ep y' (\sigma) , \;
{\partial\over \ep \partial y _{n}}
\sim {\delta \over \delta y (\sigma )}.
\eeq
At first glance it seems that the differential operators which correspond to
multi-commutators
of the matrix fields
are suppressed by some powers of $\ep$ in the continuum limit.
However we cannot simply say that they do not contribute in the continuum
theory because they can renormalize the operators which survive 
in the naive continuum limit. 
We may draw an analogy with the
quantum field theory on the lattice here.
The lattice action may be expanded formally in terms of the lattice
spacing $a$.
Although the operators which are suppressed by the powers of $a$
formally vanish in the continuum, we cannot simply neglect them
because they may renormalize the relevant operators.
In fact we can write down many lattice actions which possess the identical
continuum limit.
Although they are not unique, the continuum limit
is universal which enables us to define
a unique continuum theory.
We assume a similar universality on the differential operators
in the loop space.  Namely we expect that the differential operators we consider
in the subsequent sections represent
unique quantities in the continuum limit although they are certainly
not unique with finite $\ep$.

\section{Loop equations}
\setcounter{equation}{0}
In this section, we often show only the naive leading 
terms in the loop equations.
As we have explained in the last section, it should be understood 
that they simply represent the universality classes 
of the differential operators.
The basic equations we consider here are the following 
Schwinger-Dyson equations:
\beq
\int dAd\psi 
{\partial \over \partial A_{\mu}^\alpha}
\{Tr[t^\alpha v(C^1)]w(C^2)w(C^3)\cdots w(C^l)exp(-S)\}=0 ,
\label{eqA}
\eeq
\beq
\int dAd\psi 
{\partial \over \partial \psi ^\alpha }
\{Tr[t^\alpha v(C^1)]w(C^2)w(C^3)\cdots w(C^l)exp(-S)\}=0 .
\label{eqpsi}
\eeq
We also need an equation which is related to 
the local reparametrization invariance
of the Wilson loop:
\beq
\{(k^{\mu}_n +k^{\mu}_{n+1})({\pa \over \ep\pa  k^{\mu}_{n+1}}
-{\pa \over \ep\pa k^{\mu}_{n}})
+(\bar{\lambda} _n +\bar{\lambda} _n)({\pa \over \ep\pa \bar{\lambda}_{n+1}}
-{\pa \over \ep\pa \bar{\lambda}_{n}})
+\cdots \}w(C) = 0 .
\label{regrep}
\eeq
It is easy to check that the above expression is satisfied up to $O(\ep )$
without the terms expressed by the dots. 
Since the $O(\ep ^2 )$ contribution is commutators of 
three matrix fields, it can be expressed as differential
operators in the loop space as it is
explained in the previous section.  
It is clear now that we can satisfy this identity
by iterating such a procedure.
This identity implies the standard local reparametrization invariance of
the Wilson loop
in the continuum limit.


By multiplying $\frac{1}{2}(k^{\mu}_{1}+k^{\mu}_{M})$ to eq.(\ref{eqA}), we find
\beqa
& & \{-{3\over 2g^2 N\ep ^2 }\Delta_M({\partial\over \ep\partial k})^2 
-i{1\over g^2N\ep}\hat{k}_{\mu M}
\Delta_M({\pa\over \ep\pa \bar{\lambda} })\Gamma^0\Gamma^{\mu}
{\pa\over \ep\pa \bar{\lambda}_M}\}<w(C^1)w(C^2)\cdots w(C^l)>\n
&& +{1\over N}\sum_{j=1}^{M}\hat{k}_{\mu M}\hat{k}_{j}^{\mu}
<Tr[\prod _{n=j+1}^M U_n]
Tr[\prod _{m=1}^j U_m]
w(C^2)\cdots w(C^l)>\n
&&+{1\over N}\sum_{b=2}^l\sum_{j=1}^{M^{b}}\hat{k}_{\mu M}\hat{k}_{j}^{\mu b}
<Tr[\prod_{n=j+1}^{M^b}U^b_n v(C^1)\prod_{m=1}^{j}U^b_m]w(C^2)\cdots
\check{w}(C^b)
\cdots w(C^l)> \n
&=&0 ,
\label{covloop}
\eeqa
where $\check{w}(C^b)$ implies the absence of the Wilson loop $w(C^b)$,
and we have also introduced the notations
$\Delta_n(y)\equiv y_{n+1}-y_n$ and $\hat{y}_n\equiv {1\over 2}(y_n+y_{n+1})$.

On the other hand from eq.(\ref{eqpsi}), we obtain
\beqa
&&{i\over g^2N\ep}\Delta_M({\pa \over \ep\pa k^{\mu}})\Gamma ^{\mu}
({\pa \over \ep \pa \bar{\lambda}_M}+{\pa \over \ep \pa \bar{\lambda}_1})
<w(C^1)w(C^2)\cdots w(C^l)>\n
&&+{1\over N}\sum_{j=1}^M\hat{\lambda}_j
<Tr[\prod _{n=j+1}^M U_n]
Tr[\prod _{m=1}^j U_m]
w(C^2)\cdots w(C^l)>\n
&&+{1\over N}\sum_{b=2}^l\sum_{j=1}^{M^{b}}\hat{\lambda }_j^b
<Tr[\prod_{n=j+1}^{M^b}U^b_n v(C^1)\prod_{m=1}^{j}U^b_m]w(C^2)\cdots
\check{w}(C^b)
\cdots w(C^l)> \n
\label{psiloop}
&=&0 .
\eeqa

\begin{figure}
\begin{center}
\leavevmode
\epsfbox{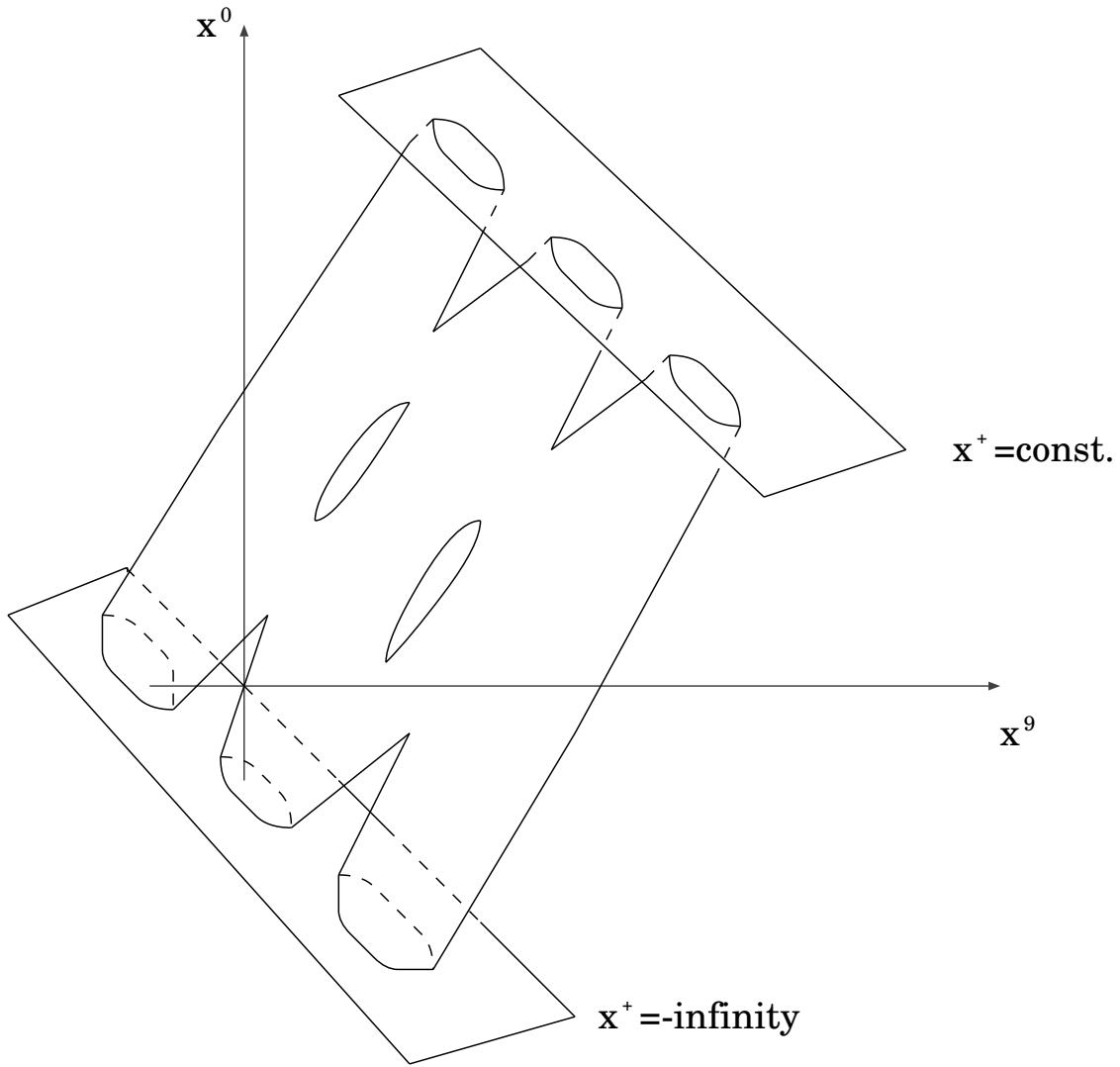}
\caption{The setting of our problem. We consider the configurations of
the Wilson loops which possess the identical light-cone time $x^{+}$. 
We also consider a group of the Wilson loops at $x^{+}=-\infty$ which represent
a particular initial state.}
\label{lightcone}
\end{center}
\vspace{1cm}
\end{figure}

In order to make the connections with the light-cone string field theory,
we consider the configurations of the Wilson loops which possess the identical
light-cone time $x^+$.  Namely we perform the Fourier transformations of the
Wilson loop from $k^-_n$ to $x^{+}_{n}$ and consider such configurations that
$x^{+}_{n}=x^{+}$ for all the strings.
We may also consider a group of the Wilson loops
at $x^+ =-\infty$ which represent a particular initial state. 
Strings which are created by the Wilson loops
at $x^+=-\infty$ evolve in time and after splitting and joining
they are eventually terminated by the Wilson loops at $x^+$.
This is the setting of our  problem which is illustrated in Fig. \ref{lightcone}.
Our strategy is to derive the light-cone Hamiltonian which
governs the evolution of the strings from the loop equation.
We can also put $k^+_n=1$ on the light-cone
by using the reparametrization invariance of the Wilson loop. 
The fermionic variables $\lambda$ can be decomposed into $8_s$ and $8_c$
representations of $SO(8)$ as $\lambda _a$ and $\lambda _{\da}$.
We eliminate $\lambda _{\da}$ and its conjugate $\theta_{\da}$
by using the loop equation
(\ref{psiloop}) just like 
eliminating the half of the fermionic degrees of freedom in the light-cone
field theory
by using the equation of motion.

Our matrix model action can be rewritten as follows in the $SO(8)$
decomposition:
\beqa
S & = &
-{1\over g^2} Tr({1\over 4}[A_{\mu},A_{\nu}][A^{\mu},A^{\nu}] \n
& & +{1\over 2}(\sqrt{2}\psi _{{a}}[A^+ ,\psi _{{a}}]
-\psi_{a}\gamma ^i_{{a}\da}[A_i,\psi _{\da}] 
-\psi_{\da}\gamma ^i_{\da{a}}[A^{i},\psi _{{a}}]
+\sqrt{2}\psi _{\da}[A^- ,\psi _{\da}]) ).
\label{SO(8)decomposition}
\eeqa
We recall that the coupling constant of this model $g$ 
possesses the dimension of length squared since we identify the
diagonal elements of $A_{\mu}$ as the spacetime coordinate in the
semiclassical region.

In the light-cone setting we just explained, 
the loop equation (\ref{covloop}) becomes as follows:
\beqa
&&\hat{k}^-_M<w(C^1)w(C^2)\cdots w(C^l)>\n
&=& Z^{-1}\{(
(\hat{k}^i_M)^2+{3\over 2g^2N\ep^2}(\Delta_M(x^i))^2
-i{\sqrt{2}\over g^2N}{\Delta_M(\th _a)\over \ep}\hat{\th_a}_M\n
&&
-{i\over g^2N\ep}\hat{k}^i_M\gamma^i_{a\da}
(\Delta_M(\th _{\da})\hat{\th_{a}}_M+\Delta_M(\th _{a})\hat{\th_{\da}}_M)
)<w(C^1)w(C^2)\cdots w(C^l)>\n
&&
+{1\over N}\sum_{j=1}^{M-1}\hat{k}_{\mu M}\hat{k}^{\mu}_j
<Tr[\prod _{n=j+1}^M U_n]
Tr[\prod _{m=1}^j U_m]
w(C^2)\cdots w(C^l)>\n
&&+{1\over N}\sum_{b=2}^l\sum_{j=1}^{M^{b}}\hat{k}_{\mu M}\hat{k}_{j}^{\mu b}
<Tr[\prod_{n=j+1}^{M^b}U^b_n v(C^1)\prod_{m=1}^{j}U^b_m]w(C^2)\cdots
\check{w}(C^b)
\cdots w(C^l)> \},\n
\label{regloop}
\eeqa
where $Z=2(1+{i\over \sqrt{2}g^2N\ep}\Delta_{M}(\th_{\da})\hat{\th}_{\da M})$.
We have also introduced conjugate variables
$x^{\mu}_{n}= i{\pa \over\pa \ep k_{n\mu}}$ and 
$\th_{n} \equiv {\pa\over \pa \ep\la_{n}}$.

In this equation, we have separated the minimal string contribution
$(j=M)$ from the 
third line of the right-hand side of the equation which represents the splitting
of the string and it is combined into the
free Hamiltonian of the string. The string of vanishing length has
$w(C)=Tr\,1=N$. 
The Wilson loop of a string with finite length cannot have the magnitude of
$O(N)$
because of the conservation of 
the light-cone momentum $p^+$  and the  
large-N factorization. In fact the loop equation (\ref{regloop})
implies
that its magnitude is of $O(1)$.

Since we put $k^+_n=1$, the total length of the 
string should be proportional to the total light-cone momentum $p^+=M\ep$.
We note that the string coupling constant which governs the strength
of the splitting and joining of strings is proportional to $1/N$
as expected. The string interaction conserves the bosonic and fermionic 
momentum densities $k_n$ and $\la_n$ locally.
If we let $N\rightarrow \infty$ in eq.(\ref{regloop}) before taking
$\ep \rightarrow 0$, the interaction terms can be neglected.
Hence we find that a free string theory is obtained in the
large-N limit.

We can eliminate $\lambda _{\da}$ and $\th _{\da}$ from eq.(\ref{regloop}) 
iteratively by 
using the following loop equations which are obtained from
eq.(\ref{psiloop}): 
\beqa
&&\hat{\th}_{\da M}<w(C^1)w(C^2)\cdots w(C^l)>\n
&=&{g^2N\over 2}({\Delta_M(x^i )\over \ep}\gamma^i_{a\da})^{-1}
\{\hat{\la}_{a M}<w(C^1)w(C^2)\cdots w(C^l)>\n
&&+{1\over N}\sum_{j=1}^{M-1}\hat{\la}_{aj}
<Tr[\prod _{n=j+1}^M U_n]
Tr[\prod _{m=1}^j U_m]
w(C^2)\cdots w(C^l)>\n
&&
+{1\over N}\sum_{b=2}^l\sum_{j=1}^{M^b}\hat{\la}_{aj}^b
<Tr[\prod_{n=j+1}^{M^b}U^b_n v(C^1)\prod_{m=1}^{j}U^b_m]w(C^2)\cdots
\check{w}(C^b)
\cdots w(C^l)>\} ,
\label{thda}
\eeqa
and
\beqa
&&\hat{\la}_{\da M}<w(C^1)w(C^2)\cdots w(C^l)>\n
&=&({2\sqrt{2}\over g^2N}{\Delta_M(x^- )\over \ep}\hat{\th}_{\da M}
+{2\over g^2N}{\Delta_M(x^i )\over \ep}\gamma^i_{\da a}
\hat{\th}_{a M})<w(C^1)w(C^2)\cdots w(C^l)>\n
&&-{1\over N}\sum_{j=1}^{M-1}\hat{\la}_{\da j}
<Tr[\prod _{n=j+1}^M U_n]
Tr[\prod _{m=1}^j U_m]
w(C^2)\cdots w(C^l)>\n
&&
-{1\over N}\sum_{b=2}^l\sum_{j=1}^{M^b}\hat{\la}_{\da j}^b
<Tr[\prod_{n=j+1}^{M^b}U^b_n v(C^1)\prod_{m=1}^{j}U^b_m]w(C^2)\cdots
\check{w}(C^b)
\cdots w(C^l)> .
\label{lada}
\eeqa

Let us concentrate on the free propagation part of the string first.
Although the large-N limit of the loop equation (\ref{regloop}) 
contains not only quadratic terms but also nonlinear
terms, it is reduced to the following simple Hamiltonian in the continuum limit: 
\beq
p^-= \int _0^{p^+} d\sigma {1\over 2}
\{c k(\sigma )^i k(\sigma )^i
+c'\beta  (x'_i (\sigma))^2 
-ic''{\beta}\theta _{a}'(\sigma)
\theta _{a}(\sigma)
-ic'''\lambda_{a}(\sigma )'\lambda_{a}(\sigma )\} ,
\label{freeHam}
\eeq
where $\beta \equiv 1/g^2N$.
It is because the only effect of the nonlinear operators in the
continuum limit is the renormalization of the coefficients
of the quadratic operators since they are suppressed by the powers of $\ep$.
In particular $\la _a '\la _a$ term is generated by the renormalization effect
after eliminating $\th _{\da}$ variables as it will be demonstrated shortly.
As it will be shown in the next section,
the form of the free light-cone Hamiltonian in the loop space is fixed uniquely
by the power counting and the $N=2$ supersymmtery. 

This Hamiltonian is identical to that of type IIB superstring theory\cite{BGS} 
if $\lambda_{a}(\sigma )$ and $\th _{a}(\sigma )$ are rescaled 
appropriately and rotated 
by a complex phase factor $\eta = exp({i\pi\over 4})$ as follows:
\beq
\la _{a}(\sigma ) \rightarrow \eta \la _{a}(\sigma ) ,
~\th _{a}(\sigma ) \rightarrow \eta^*  \th _{a}(\sigma ) .
\label{phase}
\eeq
We elaborate more on this point in connection with the supercharges 
in the next section.

The string tension of this action eq.(\ref{freeHam})
is found to be $(\alpha ')^2 \sim 1/\beta = g^2 N$
which is again the standard combination to be held fixed in the large-N limit.
Since $g$ has the dimension of the square of length, such an identification is
consistent.
Therefore we have obtained the standard light-cone Hamiltonian
of free type IIB superstring in the large-N limit. 
In particular our type IIB matrix model
is shown to possess the identical spectrum with free type IIB superstring
in the large-N limit.

We now explain how the nonlinear operators renormalize
the quadratic free string propagator by concrete examples.
Since they involve $\th_{\da}$, we need to eliminate it
iteratively by using eq.(\ref{thda}).
Let us consider the naive leading contribution in eq.(\ref{thda}):
\beq
\th _{\da}<w(C)>
={x_i ' (\sigma )\gamma ^i_{\da a}\over (x_i ' (\sigma ))^2}
{g^2N\over 2}\la _{a}(\sigma )<w(C)>.
\label{eqtha}
\eeq
We assume first that a quadratic Hamiltonian such as eq.(\ref{freeHam}) correctly
describes the free propagation of the Wilson loops of the matrix
model. This assumption can be justified by showing that the nonlinear
terms are indeed negligible in the continuum limit except for finite 
renormalization of the quadratic terms.
Since we are dealing with free two dimensional field theories,
we can utilize standard techniques of conformal field theory
to estimate the effects of the nonlinear terms.
We note that the $(x_i ' (\sigma ))^2$ is of $1/\ep ^2$ since
we have a cutoff length $\ep$.
So we may expand $(x_i ' (\sigma ))^2 = 1/ \alpha \ep^2 +:(x_i ' (\sigma ))^2 :$
where $:y:$ denotes the normal ordered operator constructed out of $y$. 
$\alpha$ in the denominator is a quantity proportional to $\sqrt{\beta}$ on the
dimensional grounds.

In this way, the left-hand side of the eq.(\ref{eqtha}) becomes
\beq
\ep^2{\alpha \over {2}\beta}
x_i ' (\sigma )\gamma ^i_{\da a}(1-\ep^2\alpha :(x_i ' (\sigma ))^2 :
+\cdots )\la _{a}(\sigma )<W(C)>.
\eeq
In principle we can eliminate $\th_{\da} (\sigma )$  in eq.(\ref{regloop}) 
through this procedure, and we obtain many terms with various powers of $\ep$.
However as we discuss in the next section the powers of $\epsilon$ 
can be understood in a simple dimensional analysis. 
We can assign the standard two dimensional 
canonical dimensions to the operators. The quadratic part of the light-cone
Hamiltonian
has the canonical dimension of two, and as we will see in the next section, 
operators with lower dimensions do not appear because of the $N=2$ supersymmetry.
Therefore we can show that the nonlinear
terms of the light-cone Hamiltonian is indeed irrelevant in the continuum limit
except for finite renormalizations of the quadratic terms.

Finally we consider the string interactions. In principle we can evaluate
the Hamiltonian $k^{-}_{n}$ by iteratively using eqs.(\ref{regrep}),
(\ref{regloop}), (\ref{thda}) and (\ref{lada}) to eliminate
the operators 
$\frac{x^{-}_{n+1}-x^{-}_{n}}{\ep}$,
$\theta_{n\da}$ and $\lambda_{n\da}$ in the right-hand side 
of eq.(\ref{regloop}). 
In this procedure, various interaction terms of
order $1/N^k$ are generated, which represent processes where
the $k+2$ strings interact at one point {\it i.e.} $(k+2)-$Reggeon vertices. 
However we will discuss in the next section that
these interactions are completely controllable again by an analysis based on the
symmetries and a power counting of $\ep$ at the interaction points.

\section{N=2 supersymmetry and general proof}
\setcounter{equation}{0}
In this section, we give a general proof of our assertion in the
previous sections by using a power counting and a symmetry analysis 
based on the $N=2$
supersymmetry, $SO(8)$ invariance and the parity symmetry on the string 
worldsheet.

\subsection{Power counting and parity symmetry}
In order to perform a power counting for $\ep$, we first introduce 
a mass dimension on the worldsheet through 
the relation $[\ep]=-1$ and determine the dimension of each field. 
By demanding the IIB matrix model action (\ref{SO(8)decomposition}) 
to be dimensionless, we obtain
\beq
[A^i]=0, \: [A^+]=-[A^-], \: [A^+]=-2[\psi_a] \: \mbox{and} \:
                             [A^-]=-2[\psi_{\da}].
\eeq
From the definition of the Wilson loop (\ref{Wilsonloop}), we also 
read off the relations
\beqa
& &[k^{+}_{n}]+[A^-]=1, \n
& &[k^{-}_{n}]+[A^+]=1, \n
& &[k^{i}_{n}]+[A^i]=1, \n
& &[\lambda_{na}]+[\psi_a]=1, \n
& &[\lambda_{n\da}]+[\psi_{\da}]=1.
\eeqa
Noting that we should set $[k^{+}_{n}]$ to be zero since $k^{+}_{n}=1$ 
in our light-cone setting, we can determine 
the dimensions of all quantities as follows:
\beqa
& &[k^{+}_{n}]=0,\: [k^{-}_{n}]=2,\: [k^{i}_{n}]=1, \:
[\lambda_{na}]=\frac{1}{2}, \: [\lambda_{n\da}]=\frac{3}{2}, \n
& &[x^{+}_{n}]=-1,\: [x^{-}_{n}]=1,\: [x^{i}_{n}]=0, \:
[\theta_{na}]=\frac{1}{2}, \: [\theta_{n\da}]=-\frac{1}{2}, \n
& &[A^+]=-1, \: [A^-]=1, \: [A^i]=0, \: [\psi_a]=\frac{1}{2},
\: [\psi_{\da}]=-\frac{1}{2},
\label{dimension}
\eeqa
where $x^{\mu}_{n}=i\frac{1}{\ep}\frac{\pa}{\pa k_{n\mu}}$ and
$\theta_n=\frac{1}{\ep}\frac{\pa}{\pa \lambda_n}$.

Next we define a symmetry which corresponds to the parity
on the string worldsheet. It is seen easily
that the IIB matrix model action (\ref{action}) is formally invariant 
under the following
transformation:
\beqa
& &A_{\mu} \rightarrow A_{\mu}^{t}, \n 
& &\psi \rightarrow -i\psi^t.
\label{paritytransfmatrix}
\eeqa
This transformation flips the direction of the Wilson loop in the following way:
\beq
w(C)=Tr(P\prod^{M}_{n=1}e^{i\ep(k^{\mu}_{n}A_{\mu}+\bar{\lambda}_n \psi)})
\rightarrow
Tr(P\prod^{M}_{n=1}e^{i\ep(k^{\mu}_{M+1-n}A_{\mu}
                            -i\bar{\lambda}_{M+1-n} \psi)}).
\eeq
Therefore our theory has a symmetry under the transformation
\beqa
& &k^{\mu}_{n} \rightarrow k^{\mu}_{M+1-n}, \n
& &\lambda_{n} \rightarrow i\lambda_{M+1-n},
\label{paritytransffield1}
\eeqa
which we identify with the worldsheet parity. We also obtain 
the parity transformation for the dual variables $x^{\mu}_{n}$ and 
$\theta_n$:
\beqa
& &x^{\mu}_{n} \rightarrow x^{\mu}_{M+1-n}, \n
& &\theta_{n} \rightarrow -i\theta_{M+1-n}.
\label{paritytransffield2}
\eeqa

\subsection{$N=2$ supersymmetry}
As is discussed in \cite{IKKT}, the IIB matrix model possesses 
the $N=2$ supersymmetry:
\beqa
\delta^{(1)} \psi &=& \frac{i}{2}[A_{\mu},A_{\nu}]\Gamma^{\mu\nu}\varepsilon, \n
\delta^{(1)} A_{\mu} &=& i\bar{\varepsilon}\Gamma_{\mu}\psi, 
\label{N=2SUSY1}
\eeqa
and 
\beqa
\delta^{(2)}\psi&=&\sqrt{2}\xi, \n
\delta^{(2)}A_{\mu}&=&0.
\label{N=2SUSY2}
\eeqa
We can determine the dimensions and parities of the parameters $\varepsilon$ 
and $\xi$ by comparing the both sides of eqs.(\ref{N=2SUSY1}) and 
(\ref{N=2SUSY2}) respectively, 
\beqa
& &[\varepsilon_a]=[\xi_a]=\frac{1}{2}, \:
[\varepsilon_{\da}]=[\xi_{\da}]=-\frac{1}{2}, \n
& &\varepsilon_a \rightarrow i\varepsilon_a, \:
   \varepsilon_{\da} \rightarrow i\varepsilon_{\da},\n
& &\xi_a \rightarrow -i\xi_a, \:
   \xi_{\da} \rightarrow -i\xi_{\da}.
\label{SUSYparameter}
\eeqa
This fixes the dimensions and parities of the supercharges $Q^1$ and $Q^2$ 
since $\varepsilon_a Q^{1}_{a}+\varepsilon_{\da} Q^{1}_{\da}
+\xi_a Q^{2}_{a}+\xi_{\da} Q^{2}_{\da}$ generates the transformations
(\ref{N=2SUSY1}) and (\ref{N=2SUSY2}):
\beq
[Q^{1}_{a}]=[Q^{2}_{a}]=-\frac{1}{2}, \:
[Q^{1}_{\da}]=[Q^{2}_{\da}]=\frac{1}{2}, 
\label{superchargedimension}
\eeq
\beqa
& &Q^{1}_{a} \rightarrow -iQ^{1}_{a}, \:
   Q^{1}_{\da} \rightarrow -iQ^{1}_{\da},\n
& &Q^{2}_{a} \rightarrow iQ^{2}_{a}, \:
   Q^{2}_{\da} \rightarrow iQ^{2}_{\da}.
\label{superchargeparity}
\eeqa
Here we note that eqs.(\ref{superchargedimension}) and 
(\ref{superchargeparity})
are consistent with the anti-commutation relations
\beqa
& &\{Q^1,Q^1\}=0, \: \{Q^2,Q^2\}=0, \n
& &\{Q^{1}_{a},Q^{2}_{b}\}=2P^+\delta_{ab}, \n
& &\{Q^{1}_{a},Q^{2}_{\da}\}=\sqrt{2}P^i\gamma^{i}_{a\da}, \: 
\{Q^{1}_{\da},Q^{2}_{a}\}=\sqrt{2}P^i\gamma^{i}_{\da a}, \n 
& &\{Q^{1}_{\da},Q^{2}_{\dot{b}}\}=2H\delta_{\da\dot{b}}.
\label{SUSYanticom}
\eeqa

\subsection{Free parts of supercharges and Hamiltonian}
The supercharges $Q^1$ and $Q^2$ can be expressed as differential operators
on the loop space using the Ward identities. In principle we can 
eliminate the operators 
$k^{-}_{n}$, $\frac{x^{-}_{n+1}-x^{-}_{n}}{\ep}$,
$\lambda_{n\da}$ and $\theta_{n\da}$ by repeatedly using
the loop equations and the reparametrization invariance as is discussed for
$k^{-}_{n}$ in the previous section. Note that we obtain interaction terms 
through this procedure. 
However as we will see just below, the
forms of their continuum limit are completely determined by the dimension,
parity and $SO(8)$ invariance.
First we concentrate on free parts of the supercharges $Q^1$ and $Q^2$
{\it i.e.} consider only the leading contribution of the $1/N$ expansion.
By using the power counting, eqs.(\ref{dimension}) and 
(\ref{superchargedimension}), 
$SO(8)$ invariance and the parity symmetry, eqs.(\ref{paritytransffield1}), 
(\ref{paritytransffield2}) and
(\ref{superchargeparity}), we can deduce the following
forms of free supercharges in the $\ep \rightarrow 0$ limit:
\beqa
& &Q^{1}_{free \, a}=\int d\sigma a_1 \theta_a(\sigma), \n
& &Q^{2}_{free \, a}=\int d\sigma a_2 \lambda_a(\sigma), \n
& &Q^{1}_{free \, \da}=\int d\sigma 
                       (b_1x'^i(\sigma)\gamma^{i}_{\da a}\lambda_a(\sigma)
                       +c_1k^i(\sigma)\gamma^{i}_{\da a}\theta_a(\sigma)), \n
& &Q^{2}_{free \, \da}=\int d\sigma 
                       (b_2k^i(\sigma)\gamma^{i}_{\da a}\lambda_a(\sigma)
                       +c_2x'^i(\sigma)\gamma^{i}_{\da a}\theta_a(\sigma)).
\label{freesupercharge}
\eeqa
Note that the integration  possesses the dimension $-1$ since 
$\int d\sigma=\ep\sum_n$. In eq.(\ref{freesupercharge}) we have excluded
terms such as $\frac{1}{\ep}x^i\gamma^i\lambda$ by translation invariance.
It is easy to see that all possible terms which appear with negative powers
of $\ep$ are forbidden by the symmetries. In this sense the existence of the 
continuum limit is guaranteed by the symmetries. We can also fix 
undetermined coefficients in eq.(\ref{freesupercharge})
by the $N=2$ supersymmetry (\ref{SUSYanticom}) as follows. 
From $\{Q^{1}_{a},Q^{2}_{b}\}=2P^+\delta_{ab}$, 
$\{Q^{1}_{a},Q^{2}_{\da}\}=\sqrt{2}P^i\gamma^{i}_{a\da}$ and 
$\{Q^{1}_{\da},Q^{2}_{a}\}=\sqrt{2}P^i\gamma^{i}_{\da a}$, we obtain
\beq
a_1a_2=2, \: a_1b_2=\sqrt{2} \: \mbox{and} \: a_2c_1=\sqrt{2}.
\eeq
Therefore eq.(\ref{freesupercharge}) is reduced to
\beqa
& &Q^{1}_{free \, a}=a_1\int d\sigma \theta_a(\sigma), \n
& &Q^{2}_{free \, a}=\frac{2}{a_1}\int d\sigma \lambda_a(\sigma), \n
& &Q^{1}_{free \, \da}=\int d\sigma 
               (\frac{a_1}{\sqrt{2}}k^i(\sigma)\gamma^{i}_{\da a}\theta_a(\sigma)
                  +b_1x'^i(\sigma)\gamma^{i}_{\da a}\lambda_a(\sigma)), \n
& &Q^{2}_{free \, \da}=\int d\sigma 
             (\frac{\sqrt{2}}{a_1}k^i(\sigma)\gamma^{i}_{\da a}\lambda_a(\sigma)
                       +c_2x'^i(\sigma)\gamma^{i}_{\da a}\theta_a(\sigma)).
\label{freesuperchargefinal}
\eeqa
The free part of the Hamiltonian $\sum k^{-}_{n}$ is obtained by 
$\{Q^{1}_{\da},Q^{2}_{\dot{b}}\}=2H\delta_{\da\dot{b}}$ as
\beq
H_{free}=\int d\sigma (\frac{1}{2}k^i(\sigma)^2+\frac{1}{2}b_1c_2x'^i(\sigma)^2
-i\frac{b_1}{\sqrt{2}a_1}\lambda'_a(\sigma)\lambda_a(\sigma)
           -i\frac{a_1c_2}{2\sqrt{2}}c_2\theta'_a(\sigma)\theta_a(\sigma)),
\label{freehamiltonian}
\eeq
In order to compare these results with the Green-Schwarz light-cone formalism,
we redefine the fermionic variables as
\beq
\lambda_a = \sqrt{\frac{a_1}{b_1}} \eta \check{\lambda}_a \:
\mbox{and} \: \theta_a = \sqrt{\frac{b_1}{a_1}} \eta^* \check{\theta}_a,
\label{fermionrescaling}
\eeq
where $\eta=e^{\frac{\pi i}{4}}$. 
We also introduce rescaled supercharges $\check{Q}^1$ and $\check{Q}^2$ by
\beq
Q^{1} = \sqrt{a_1b_1} \eta^{*} \check{Q}^{1} \: \mbox{and} \:
Q^{2} = \frac{\eta}{\sqrt{a_1b_1}} \check{Q}^{2}.
\label{superchargerescaling}
\eeq
In terms of these new quantities (\ref{freesuperchargefinal}) 
and (\ref{freehamiltonian}) become
\beqa
& &\check{Q}^{1}_{free \, a}=\int d\sigma \check{\theta}_a(\sigma), \n
& &\check{Q}^{2}_{free \, a}=2\int d\sigma \check{\lambda}_a(\sigma), \n
& &\check{Q}^{1}_{free \, \da}=\int d\sigma 
        (\frac{1}{\sqrt{2}}k^i(\sigma)\gamma^{i}_{\da a}\check{\theta}_a(\sigma)
                  +ix'^i(\sigma)\gamma^{i}_{\da a}\check{\lambda}_a(\sigma)), \n
& &\check{Q}^{2}_{free \, \da}=\int d\sigma 
             (\sqrt{2}k^i(\sigma)\gamma^{i}_{\da a}\check{\lambda}_a(\sigma)
              -ib_1c_2x'^i(\sigma)\gamma^{i}_{\da a}\check{\theta}_a(\sigma)), \n
& &H_{free}=\int d\sigma (\frac{1}{2}k^i(\sigma)^2
                           +\frac{1}{2}b_1c_2x'^i(\sigma)^2
+\frac{1}{\sqrt{2}}\check{\lambda}'_a(\sigma)\check{\lambda}_a(\sigma)
    -\frac{1}{2\sqrt{2}}b_1c_2\check{\theta}'_a(\sigma)\check{\theta}_a(\sigma)),
\eeqa
which completely agree with the light-cone Green-Schwarz free Hamiltonian 
and supercharges for type IIB superstring. This fact also justifies the analytic 
continuation introduced for fermionic fields in ref. \cite{IKKT}.
We also note that we have obtained the relation $b_1c_2 \sim 1/\alpha'^2$,
and $b_1c_2$ should be equal to $1/g^2N$ multiplied by some numerical constant
as is illustrated in the previous section. 

\subsection{Interaction parts of supercharges and Hamiltonian}
In this subsection, we examine the structure of the interaction parts of 
the supercharges and the Hamiltonian. First we consider the contributions
of order $1/N$, which correspond to $3-$Reggeon vertices in string field theory.
Since our free Hamiltonian is equal to that of the Green-Schwarz light-cone
formalism, we can use the same arguments as in light-cone string field theory.
In general, the operators inserted near the interaction points in $3-$Reggeon
vertices generate divergences coming from the Mandelstam mapping.
Since our Wilson loops are written by the variables $k^i$ and $\lambda$,
the corresponding $3-$Reggeon vertices should consist of delta functions 
representing the matching of three strings in the $k-\lambda$ space, 
which is the same as in ref. \cite{BGS}.
Therefore the $k^i$, $x'^i$ and $\lambda_a$ diverge as $1/\sqrt{\ep}$ near
the interaction points while $\theta_a$ is of order $\ep^0$ there.
We also note that every derivative of $\sigma$ acting on the fields introduces
an extra factor of $1/\sqrt{\ep}$.
Therefore the interaction part at order $1/N$ of the supercharges possesses 
the following general structure:
\beq
\frac{1}{N\ep}\int d\sigma \int d\sigma_1 \ep^{\eta}(k^i)^{\alpha} (x'^i)^{\beta}
(\lambda_a)^{\gamma}(\theta_a)^{\delta}(\mbox{derivative})^{\kappa}
(\mbox{products of delta functions for }k^i \mbox{ and } \lambda_a),
\eeq
where $k^i$, $x'^i$, $\lambda_a$ and $\theta_a$ represent the operators inserted
near the interaction points, and $\kappa$ is the total number of derivatives 
acting on these operators. Note that we have extracted the factor $1/\ep$ when 
we rewrite the sum for the interaction points to the integral over $\sigma_1$.

For example, let us consider the interaction part of $Q^{1}_{\da}$. 
In this case, the dimensional analysis 
$[Q^{1}_{int \, \da}]=[Q^{1}_{free \,\da}]$ leads to
$-\eta+\alpha+\beta+\frac{1}{2}\gamma+\frac{1}{2}\delta
+\kappa-1=\frac{1}{2}$. 
Therefore 
the total powers of $\ep$ which appear in the interaction part of $Q^{1}_{\da}$
is evaluated as
\beqa
\zeta&=&-1+\eta-\frac{1}{2}\alpha-\frac{1}{2}\beta-\frac{1}{2}\gamma
                                -\frac{1}{2}\kappa \n
     &=&\frac{1}{2}\alpha+\frac{1}{2}\beta+\frac{1}{2}\delta+\frac{1}{2}\kappa
                                -\frac{5}{2}.
\eeqa
The case in which $\alpha=\beta=\delta=\kappa=0$ is excluded by
$SO(8)$ invariance. We can consider four cases in which $\zeta=-2$:
(1)$\alpha=1$ and $\beta=\delta=\kappa=0$, (2)$\beta=1$ and
$\alpha=\delta=\kappa=0$, (3)$\delta=1$ and $\alpha=\beta=\kappa=0$ and
(4)$\kappa=1$ and $\alpha=\beta=\delta=0$. The cases (3) and (4) are not 
permitted by $SO(8)$ invariance. If we take the large-N limit with $N\ep^2$
kept fixed, the cases (1) and (2) survive in the $\ep \rightarrow 0$ limit. 
Note that in this limit all of the other cases vanish because $\zeta$ is larger 
than $-2$ for them.
Furthermore we can restrict the values of
$\gamma$ by the parity symmetry and deduce the structure
of $Q^{1}_{int \,\da}$ as follows:
\beqa
& &Q^{1}_{int \,\da}=\frac{1}{N\ep^2}\int d\sigma \int d\sigma_1
(\sqrt{\ep}k^i((\sqrt{\ep}\lambda_a)^3+(\sqrt{\ep}\lambda_a)^7)
+\sqrt{\ep}x'^i(\sqrt{\ep}\lambda_a+(\sqrt{\ep}\lambda_a)^5)) \n
& & \hspace{4cm}
(\mbox{products of delta functions for }k^i \mbox{ and } \lambda_a).
\label{Q1intda}
\eeqa
This structure agrees with that of the light-cone 
string field theory \cite{BGS}. Applying a similar analysis to
$Q^{2}_{int \,\da}$, we obtain 
\beqa
& &Q^{2}_{int \,\da}=\frac{1}{N\ep^2}\int d\sigma \int d\sigma_
(\sqrt{\ep}k^i(\sqrt{\ep}\lambda_a+(\sqrt{\ep}\lambda_a)^5)
+\sqrt{\ep}x'^i((\sqrt{\ep}\lambda_a)^3+(\sqrt{\ep}\lambda_a)^7)) \n
& & \hspace{4cm} 
(\mbox{products of delta functions for }k^i \mbox{ and } \lambda_a)
\label{Q2intda}
\eeqa
which also agrees with the light-cone string field theory.
As for $Q^{1}_{a}$ and $Q^{2}_{a}$, no $1/N$ contribution remains non-zero 
in this limit since the minimum value of $\zeta$ is $-\frac{3}{2}$ in these
cases. Therefore we conclude that 
$Q^{1}_{int \,a}$ and $Q^{2}_{int \,a}$ are equal to zero at order $1/N$, 
which is again consistent
with the light-cone string field theory.
Note that the right-hand sides of eqs.(\ref{Q1intda}) and (\ref{Q2intda}) 
are uniquely determined
by the $N=2$ supersymmetry, as is shown in ref.\cite{BGS}.
Finally the anti-commutation relation 
$\{Q^{1}_{\da},Q^{2}_{\dot{b}}\}=2H\delta_{\da\dot{b}}$ fixes 
the interaction part of $H$, which is certainly consistent with
the light-cone string field theory.

Next we consider the contributions of order $1/N^k \: (k\geq2)$, 
which correspond to $(k+2)-$Reggeon vertices. The general
structure of the interaction part is represented as
\beqa
& &\frac{1}{N^k\ep^k}\int d\sigma \int d\sigma_1 \cdot \cdot \cdot d\sigma_{k}
       \ep^{\eta}(k^i)^{\alpha} (x'^i)^{\beta} 
       (\lambda_a)^{\gamma}(\theta_a)^{\delta}(\mbox{derivative})^{\kappa} \n
& & \hspace{5cm}
(\mbox{products of delta functions for }k^i \mbox{ and } \lambda_a).
\eeqa
From the Mandelstam mapping in these cases, it is natural to consider that
the $k^i$, $x'^i$ and $\lambda_a$ diverge as $\ep^{-\frac{k}{k+1}}$ near
the interaction points.
Therefore the total power $\zeta$ of $\ep$ is evaluated as 
\beq
\zeta=\frac{1}{k+1}(\alpha+\beta+\kappa)+\frac{1-k}{2(k+1)}\gamma
+\frac{1}{2}\delta-k-\rho,
\eeq
where $\rho$ is $\frac{1}{2}$ for $Q^{1}_{int \,a}$ and $Q^{2}_{int \,a}$, 
and $\frac{3}{2}$ for $Q^{1}_{int \,\da}$ and $Q^{2}_{int \,\da}$ and
the terms in which $\zeta\leq-2k$ survive in the $\ep \rightarrow 0$ limit
if $N\ep^2$ is fixed.
It is verified easily that there are no surviving terms for any values of $k$
in $Q^{1}_{int \,a}$ and $Q^{2}_{int \,a}$
in the $\ep \rightarrow 0$ limit, which is consistent with the light-cone string
field theory. Using $SO(8)$ invariance, we can show that 
in $Q^{1}_{int \,\da}$ and $Q^{2}_{int \,\da}$ some terms with $\gamma$ equal to
five might survive for $k=2$ and ones with $\gamma$ equal to seven for $k=2$ and
$k=3$. Presumably it is not possible to satisfy 
$N=2$ supersymmetry only by these restricted terms. 
Therefore we may conclude that there are no
contributions of order $1/N^k \: (k\geq2)$ 
in $Q^{1}_{int \,\da}$, $Q^{2}_{int \,\da}$ and the Hamiltonian, 
which is also consistent with
the light-cone string field theory.

In this way, we almost confirm that our IIB matrix model reproduces 
the light-cone string field theory for type IIB superstring. 
In particular, we have 
found the prescription of the double scaling limit in the IIB matrix model:
\beq
g^2N \sim \alpha'^2 \; \mbox{and} \; N\ep^2 \sim g_{st}^{-1}.
\eeq

\section{Conclusions and Discussions}
\setcounter{equation}{0}
In this paper, we have investigated the loop equations of the type IIB
matrix model.
We have introduced strings into the theory as the Wilson loops.
The loop equations are found to agree with the light-cone superstring field
theory
of Green, Schwarz and Brink in the double scaling limit.
What we have shown here is that the precisely the same structure naturally
emerges 
in the double scaling limit from our matrix model. 
Although we have not calculated the coefficients of the generic operators
which appear in the light-cone Hamiltonian, we have determined most of them
by using the $N=2$ supersymmetry. The remaining free parameters are
the string tension and the string coupling constant.
We are thus able to prove that the IIB matrix model indeed reproduces the
standard
perturbation series of string theory. This constitute the proof of our
previous
conjecture that our matrix model is a nonperturbative formulation of type IIB
superstring theory.
We have found that the string tension is of the order of the reduced model
coupling constant $(\alpha ')^2 \sim g^2 N$. The double scaling prescription
is to let $N \rightarrow \infty$ and $\ep \rightarrow 0$ while
$N\ep ^2$ being kept fixed. This double scaling prescription is different
from our previous estimate based on the reduced model dimensional
analysis\cite{IKKT}.

Our matrix model has been related to the type IIB Green-Schwarz superstring
action
in our previous work. In order to make this connection, we have to rotate
the phase of 
a fermionic field of Green-Schwarz action by $\pi/2$ in the complex plane. 
One of the important results of this paper is to fully justify
this analytic continuation procedure by finding the light-cone superspace
variables
through the Wilson loops.  We believe we have dispelled any suspicions on
this point.
By the way we can choose the identical or the opposite phase 
when we equate the two fermionic fields
of the type IIB Green-Schwarz superstring action
to fix the $\kappa$ symmetry.
This freedom corresponds to choose D-instantons or D-anti
instantons as our fundamental constituents of the IIB matrix model. 
Although our action does not possess the manifest symmetry between them,
we expect it to  be recovered by integrating all field configurations.
We can also construct the transformations which interchanges between them
at non-exceptional field configurations.

It has been also suggested to modify our action by adding higher dimensional
operators like Born-Infeld action\cite{olesen}. The effect of such a
modification is 
to induce higher dimensional operators in the loop space Hamiltonian.
Therefore we believe that it belongs to the same universality class
as our model.
Recall that the basic building block of our Wilson loop is the minimal
link variable 
$U _{\mu} \sim exp\{i\epsilon A_{\mu}\}$. 
We can assume here that $k_n^{\mu}$ only take integer values. 
The key element for our success to
derive our light-cone Hamiltonian is that the Wilson loops
do not possess the vacuum expectation values of order $N$ in the
large-N limit in our setting.  
Since the Wilson loop is made of the minimal links,
it is natural to cutoff the eigenvalues
of the gauge fields $A_{\mu}$ between $-\pi/\ep$ and $\pi/\ep$.
Then the expectation values of Wilson loops vanish if the eigenvalues are 
uniformly distributed. In other words we need translation invariance of 
$U(1)$ phases of  $U _{\mu}$ ($U(1) ^d$ symmetry),
which is the most crucial symmetry we have to preserve and we expect 
it is not broken spontaneously due to the supersymmetry.

Since we introduce the cutoff in the eigenvalues of $A_{\mu}$, the cutoff
also breaks supersymmetry. However the effect of the cutoff goes
away if we can take the cutoff to be infinity. Since we have shown that we can take
the continuum limit of the loop equations,  we have been consistent to
assume the
supersymmetry.
Although we have shown that the string perturbation theory follows from our
matrix model, we have largely relied on the symmetry arguments.
Therefore the precise coefficients of the string tension and the string
coupling constant is 
not determined yet.
One of our future tasks is clearly to determine them.
It is also very desirable to make nonperturbative predictions from
our matrix model since we now understand how to take the double
scaling limit. We also hope to report some progress in this respect
in the near future.


\newpage                        
\begin{thebibliography}{99}
\bibitem{MS} A. Sen, Int. J. Mod. Phys. {\bf A9} (1994) 3703;\\
J. H. Schwarz, Lett. Math. Phys. {\bf 34} (1995) 309;\\
 J. Maharana and J. Schwarz, Nucl. Phys. {\bf B390} (1993) 3.
\bibitem{Witten} E. Witten, Nucl. Phys. {\bf 443} (1995) 85.
\bibitem{Polchinski} J. Polchinski, Phys. Rev. Lett. {\bf 75} (1995) 4724.
\bibitem{BFSS}T. Banks, W. Fischler, S.H. Shenker and L. Susskind, 
{\em M Theory as a Matrix Model: a Conjecture}, hep-th/9610043.
\bibitem{IKKT}N. Ishibashi, H. Kawai, Y. Kitazawa and A. Tsuchiya,
{\em A Large-N Reduced Model as Superstring},
hep-th/9612115, to appear in Nucl. Phys. B.
\bibitem{thoot} G. 't 
Hooft, Nucl. Phys. {\bf B72} (1974) 461.
\bibitem{makeenko} Yu.M. Makeenko and A.A. Migdal, Phys. Lett. {\bf 88B}
(1979) 135.
\bibitem{RM}T. Eguchi and H. Kawai, Phys. Rev. Lett. {\bf 48} (1982) 1063.\\
      G. Parisi, Phys. Lett. {\bf 112B} (1982) 463.\\
      D. Gross and Y. Kitazawa, Nucl. Phys. {\bf B206} (1982) 440.\\
      G. Bhanot, U. Heller and H. Neuberger, Phys. Lett. {\bf 113B} (1982) 47.\\
      S. Das and S. Wadia, Phys. Lett. {\bf 117B} (1982) 228.
\bibitem{Bars} I. Bars, Phys. Lett. {\bf 245B}(1990)35.
\bibitem{Zacos} D.B. Fairlie, P. Fletcher and C.Z. Zachos,
J. Math. Phys. {\bf 31} (1990) 1088.
\bibitem{BGS} M. Green, J. Schwarz and, L. Brink, 
 Nucl. Phys. {\bf B219} (1983) 437.
\bibitem{periwal} V. Periwal, hep-th/9611103.
\bibitem{Li} M. Li, {\em Strings from IIB Matrices}, hep-th/961222.
\bibitem{chepelev} I. Chepelev, Y. Makeenko and K. Zarembo,
{\em Properties of D-branes in Matrix
Model of IIB Superstring}, hep-th/9701151.
\bibitem{smith} A. Fayyazuddin and D.J. Smith, {\em P-Brane Solutions in
IKKT IIB Matrix
Theory}, hep-th/9701168.
\bibitem{olesen} A. Fayyazuddin , Y. Makeenko, P. Olesen, D.J. Smith and K.
Zarembo,
{\em Towards a Non-perturbative Formulation of IIB Superstrings by Matrix
Models},
hep-th/9703038.
\bibitem{yoneya} T. Yoneya, {\em Schild Action and Space-Time Uncertainty
Principle
in String Theory}, hep-th/9703078.
\bibitem{olesen2} C. F. Kristjansen and P. Olesen,
{\em A Possible IIB Superstring Matrix Model with Euler Characteristic and 
a Double Scaling Limit}, hep-th/9704017.
\bibitem{GS}M. Green and J. Schwarz, Phys. Lett. {\bf 136B} (1984) 367.
\bibitem{Schild} A. Schild, Phys. Rev. {\bf D16} (1977) 1722.
\end{thebibliography}
\end{document}